# Least Squares based Estimation of Thevenin Equivalent in Noisy Distribution Grid

Taha Saeed Khan

*School of Electrical and Computer Engineering*
*Oklahoma State University, Stillwater, OK, US*
taha_saeed.khan@okstate.edu

*Abstract -* **This work presents a novel approach that synergizes the extremum seeking method with an online least squares estimation technique to accurately estimate Thevenin equivalent circuit being seen at each node in distribution grids. Thevenin's theorem offers a simplified representation of electrical networks, critical for the effective monitoring, control, and optimization of grid operations. However, real-time identification of Thevenin parameters, particularly impedance, poses significant challenges due to the dynamic nature of distribution grids. By integrating extremum seeking algorithms, which are adept at locating optima in dynamic systems without explicit model information, with the robustness of least squares estimation, we develop a novel methodology that continuously adapts to grid fluctuations. These fusion harnesses the strengths of both techniques: the extremum seeking method's non-model-based optimization capabilities and the least squares method's proficiency in estimating parameter value in a noisy environment. The result is a robust, adaptive algorithm capable of delivering reliable Thevenin parameter estimations in real-time. Our simulation results demonstrate the efficacy of the proposed method, showcasing its potential as a tool for enhanced grid management and resilience.**

## I. INTRODUCTION

The dynamic nature of power system configurations and load conditions results in constant fluctuations in grid impedance. Tracking these impedance variations is essential for several key applications within power systems. For instance, understanding changes in impedance is crucial for grid stability analysis [1], informing protective relay operation decisions [2], optimizing voltage regulation [3], detecting islanding occurrences [4], and performing accurate short circuit calculations [5].

Every method designed to estimate Thevenin parameters necessitates accurate current and voltage measurements at the Point of Common Coupling (PCC), with classification into either passive or active techniques. Passive techniques avoid the need to introduce any signal or perturbation into the grid, yet they face distinct obstacles. For example, the Extended Kalman Filter (EKF) [6] demands high-precision electronics and intricate adjustment of unknown noise parameters to yield precise outcomes. Similarly, the recursive least square (RLS) method [7] hinges on a grid model, as well as precise determination of phase and frequency for d-q reference frame transformation. Conversely, active methods involve the injection of disturbances or perturbations into the system. Compared to passive methods, they offer more straightforward Thevenin parameter estimation. Various active approaches have been put forward, ranging from those that require the injection of a single-frequency signal [8], to the introduction of voltage [9] or current impulses [10], or even the use of a full-spectrum frequency signal [11]. Despite potentially reducing complexity, these methods might necessitate intricate post-processing, significant impulse power injections, and extended estimation periods to achieve accurate results. To streamline computational demands, several researchers [12] [13] have applied Akagi's [14] strategy of modulating real and reactive power injections—known as the p-q technique—to infer grid impedance by monitoring its impact on PCC voltage.

This study introduces an innovative approach that combines extremum-seeking control with recursive least square estimation to precisely determine the Thevenin parameter of the grid seen from a node. The proposed method capitalizes on the extremum relationship between the voltage at the Point of Common Coupling (PCC) and the angle of injected current. By perturbing both the magnitude and angle of the current, the technique can effectively identify the Thevenin parameters. A significant advantage of the proposed method is its independence from grid models and the avoidance of complex estimation processes, costly hardware, and high-power impulse injections. Additionally, it exhibits robustness against external disturbances, enhancing its practicality for real-world grid analysis. This approach simplifies the Thevenin identification process while maintaining accuracy, offering a valuable tool for grid management and analysis.

In impedance estimation applications, the convergence speed of algorithms is a critical factor for multiple practical problems. For instance, applications such as stability margin calculation, voltage regulation, fault analysis etc. require impedance to be calculated as quickly as possible. Hence this work keeps the online estimation time limited to a few seconds to make it an attractive approach for online Thevenin identification.

The primary contribution of this work is the development of a method that enhances the functionality of smart inverters by introducing a dual-perturbation technique. This approach involves the persistent excitation of two distinct perturbations, one targeting the current angle and the other its magnitude, each at different frequencies. This innovative strategy facilitates the real-time identification of Thevenin parameters. By effectively leveraging the data generated during the operation of smart inverters, this method transforms it into a meaningful tool for continuous, dynamic analysis of grid characteristics. This advancement not only enables more efficient and accurate grid monitoring but also significantly contributes to the evolving field of smart grid technology.

The rest of the article is organized as follows. Section II will present the proposed technical approach and provide physical interpretation of how the combination ESC, RWLS (recursive weighted least square)/Kalman Filter method helps in identifying the Thevenin parameters. Section III will describe a simulation setup which is performed to validate the proposed technique. Section IV will discuss the results. Concluding remarks will be made in section V.

## II. TECHNICAL APPROACH

*A. Problem Formulation*:

The distribution grid is currently undergoing a significant transformation with the widespread integration of power electronic inverters, primarily to facilitate the interconnection of distributed energy resources. These inverters, beyond their primary role, also offer valuable capabilities for measurement and monitoring within the grid. Currently, widely used grid connected inverters are grid following. Grid following inverters are current source devices where the current magnitude and angle are adjusted to inject required amount of real and reactive power into the grid. These inverters can be modelled as controlled current sources injecting real and reactive power into the grid.

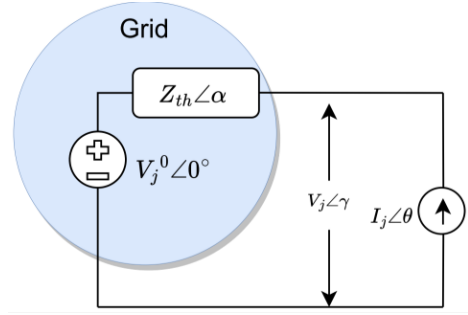

Fig. 1: An integrated inverter at node j, injecting current $I_j \angle \theta$.

A power system is essentially a complex network composed of interlinked nodes. At each of these nodes, the power system can be conceptually represented using a Thevenin equivalent circuit, which simplifies the analysis of electrical networks. When integrating an inverter, modeled as a current source, into this framework, the node to which the inverter is connected undergoes a specific transformation in its representation. This modified scenario, where a Thevenin equivalent circuit includes an inverter at a particular node (let's say node j), can be accurately depicted as shown in Fig. 1. This figure illustrates how the inverter, functioning as a current source, interacts with and influences the Thevenin equivalent circuit at the node, offering a clear and simplified view of the interaction at play within the power system at that specific point.

With the angle of Thevenin voltage ($V_j^0 \angle 0$) taken as reference, the voltage change ($\Delta V$) caused at node $j$ by injecting current ($I_j \angle \theta$) can have different values depending on the angle of current injected ($\theta$) and the angle of Thevenin impedance ($\alpha$). Hence the voltage at node $j$ can be represented as (1):

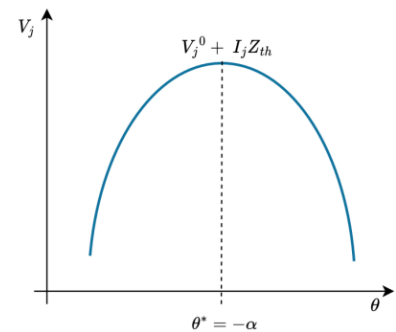

Fig. 2: An equilibrium map with an extremum for voltage at node $j$

$$V_j \angle \gamma = V_j^0 \angle 0 + I_j \angle \theta \, (Z_{th} \angle \alpha) \quad (1)$$

The square of the magnitude of this voltage can be written as:

$$(V_j)^2 = (V_j^0)^2 + (I_j Z_{th})^2 \\ + 2(V_j^0)(I_j Z_{th}) \cos(\theta + \alpha) \quad (2)$$

$$(V_j)^2 = (V_j^0 + I_j Z_{th})^2 \\ + 2(V_j^0)(I_j Z_{th})[\cos(\theta + \alpha) - 1] \quad (3)$$

$$(V_j)^2 = (V_j^0 + I_j Z_{th})^2 \\ - 4(V_j^0)(I_j Z_{th})\left[\sin^2\left(\frac{\theta + \alpha}{2}\right)\right] \quad (4)$$

It can be seen in (4) that when $\theta$ will be equal to $-\alpha$, then the voltage magnitude at node $j$ will reaches its maximum value. This can be depicted as an equilibrium map with an extremum as shown in Fig 2. In this situation $(V_j)^2 = (V_j^0 + I_j Z_{th})^2$.

In this project, ESC is proposed to find the extremum point ($\theta^* = -\alpha$) in the voltage equilibrium map. This will result in identifying the impedance angle ($\alpha$). In parallel, the magnitude of the injected current will be perturbed at a different frequency, and its impact on the voltage magnitude is observed perturbation. By feeding the observed data to the online RWLS/Kalman Filter algorithm, the magnitude of the Thevenin impedance and the Thevenin voltage are also then correctly estimated.

*B. Theoretical development:*
ESC:
Extremum Seeking Control (ESC) is a model-free type of control technique that relies on feedback to find the extremum of a static map or optimize the parameters of a dynamic system. The extremum-seeking algorithm consists of injecting perturbation into the system and observing its effect on the desired cost function. This is realized by injecting perturbation at a fixed frequency and demodulating the cost function with the same frequency at which the perturbations are applied. The observed effect is utilized to steer the system towards the optimal operating point.

Fig. 3 depicts a conventional single input-single output (SISO) ESC scheme. In a conventional ESC scheme, perturbation at a particular frequency is modulated with the applied input to the system (plant) and the output (V) is observed. The output is then passed through a high pass filter to remove any low frequency components. The effect of applied perturbation is extracted (ξ) by multiplying the filtered output with the same frequency signal as applied perturbation also called demodulation process. To extract the gradient, the demodulated signal is integrated. The operating point of the controller is then adjusted after multiplying it with an appropriate gain (K). The processes run in a closed loop and help the system to converge to an extremum while observing any real time change in the optimal point.

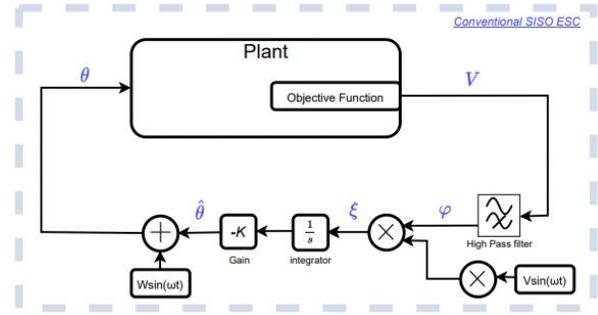

Fig. 3: Conventional Extremum seeking controller.

RWLS:
The Recursive Least Squares (RLS) algorithm with covariance form, as shown in figure 4, is an adaptive filter algorithm designed to recursively compute the least squares estimate $\theta^{(LS)}(k)$ for a system's parameters. The algorithm is used to estimate magnitude $Z^{th}$ and $V^{th}$. Figure 5 shows how perturbation and observation is carried out.

The linear model based on figure 5 can now be written as:
z = H $\theta$ + v
- z is the observed voltage $V_j$.
- H is the current $I_j$.
- $\theta$ is the Thevenin impedance $Z^{th}$.
- v corresponds to the noise/error due to imperfect filtering

The algorithm updates its estimates in real-time as new data becomes available, which makes it particularly useful for time-varying systems. The RLS algorithm with covariance form is particularly powerful because it can

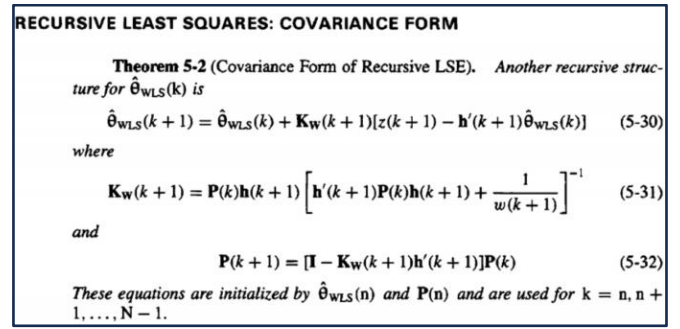

Fig. 4: Depicts the RWLS algorithm [15].

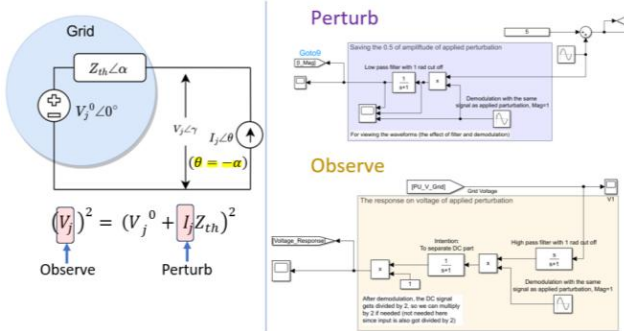

Fig. 5: Perturb and observe technique for estimating $V^{th}$ and $Z^{th}$ magnitudes [15].

adapt to changes in the system dynamics and external noise. Steps followed by the algorithm are now as follows:
- Initialization

The algorithm begins with initial estimates for the parameter vector $\theta^{(LS)}$ and the error covariance matrix $P$. The parameter vector represents the system's current estimated state, and the covariance matrix represents the uncertainty associated with this estimate.
- Prediction and Update Cycle:

At each time step k, the algorithm predicts the system's output z(k) using the current estimate of the parameter vector $\theta^{(LS)}(k)$ and the input measurement h(k). The input measurement can include current and past data points depending on the system.

Further, the prediction error is calculated by taking the difference between the actual system output z(k+1) and the predicted output $h^T(k+1)\theta^{(LS)}(k)$.
- Gain Update:

The Kalman gain Kw(k+1) is computed, which balances the prediction error with the previous estimate's uncertainty. This gain is crucial as it determines how much the new measurement will influence the updated estimate. The gain is calculated using the current error covariance matrix P(k), the input measurement h(k+1), and the noise covariance w(k+1), which accounts for the uncertainty in the measurement.
- Parameter Update:

The algorithm then updates the parameter estimate $\theta^{(LS)}(k)$ by adjusting the previous estimate based on the prediction error and the Kalman gain.
- Covariance Update:

The error covariance matrix P(k+1) is updated to reflect the new level of uncertainty after incorporating the latest measurement. This is done by reducing the previous covariance in the direction of the measurement.
- Iterative Process:

These steps are repeated in a loop for each new time step or data point, from k=n to N−1, where N is the total number of measurements or the final time step.

Bayesian Estimation and Kalman Filtering:
Another algorithm developed to estimate the $Z^{th}$ and $V^{th}$ is Kalman Filter. An introduction and implementation are provided in the appendix of this report.

III. SIMULATION

Simulation is set up in Simulink with an inverter modelled as a current source injecting current into a node in a distribution network. The current source is controlled by an extremum-seeking controller, which perturbs the angle of injected current and operates independently of the grid parameters and configuration. Another perturbation in magnitude of the current is also modulated to extract magnitude of Thevenin impedance and Thevenin voltage. Figure 6 shows the extremum seeking block developed in Simulink to inject perturbation into the current angle and estimate angle of Thevenin impedance. Figure 7 shows the RWLS block which injects perturbation into current magnitude and estimates the magnitude of Thevenin impedance and Thevenin angle.

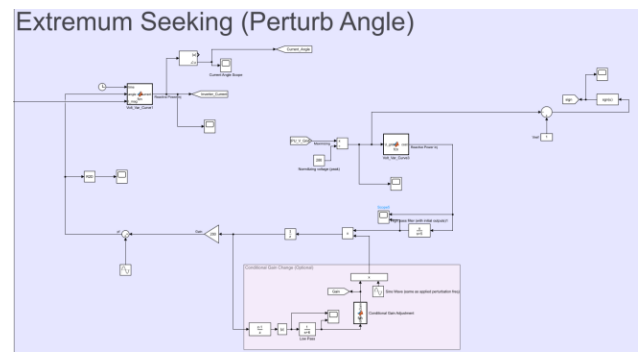

Fig. 6: ESC block implemented in Simulink environment.

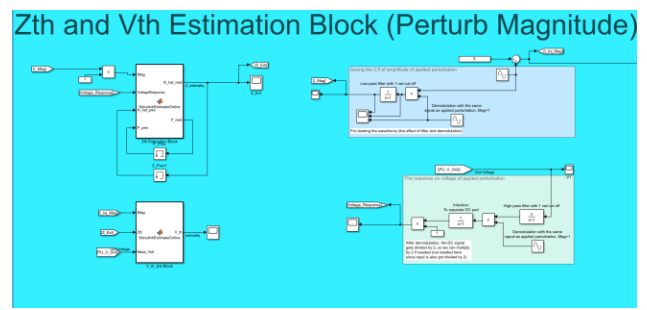

Fig. 7: ESC block implemented in Simulink environment.

During the simulation, the resistance to reactance (R/X) ratio of the circuit is manipulated over two distinct time intervals. The impedance value change are essential parameters for the simulation, reflecting changes that the system might undergo in a real-world scenario, and are critical for testing the robustness and responsiveness of the extremum-seeking control with RWLS estimation mechanism within the inverter model.

Initially, from 0 to 100 seconds, the R/X ratio is maintained at a value of 35.3243, which corresponds to an impedance angle ($\angle\alpha$) of 35.3 degrees. During this interval, the magnitude of the Thevenin impedance ($|Z^{th}|$) is set to 1.42 Ohms, and the Thevenin voltage ($V^{th}$) is held constant at 245 volts.

Subsequently, from 105 to 195 seconds, the R/X ratio of the circuit i.e. impedance angle ($\angle\alpha$) is adjusted to 54.7 degrees. In this phase, the magnitude of the Thevenin impedance ($|Z^{th}|$) is increased to 2.8 Ohms, while the Thevenin voltage ($V^{th}$) remains unchanged at 245 volts.

This variation in the corresponding impedance values in the grid and an inverter connected at node j is shown in figure 8.

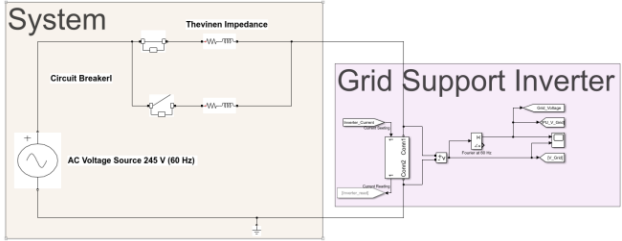

Fig. 8: ESC block implemented in Simulink environment.

## IV. RESULTS

Graph in figure 9,10 and 11 shows the performance of the proposed scheme for estimating Thevenin impedance's angle, Thevenin impedance's magnitude and Thevenin voltage respectively. The result shows how the proposed scheme correctly estimates the Thevenin circuit parameters.

## V. Conclusion

This study introduces an efficient and cost-effective method for identifying Thevenin parameters within a distribution system. Characterized by its computational simplicity and ease of implementation, this method can be seamlessly integrated into existing inverter technology. By conceptualizing the identification process as an extremum seeking problem and employing Recursive Least Square estimation, the approach systematically discerns Thevenin parameters. This dual-faceted strategy not only simplifies the computational burden but also enhances the accuracy and reliability of the parameter estimation, facilitating a more robust and intelligent grid management system.

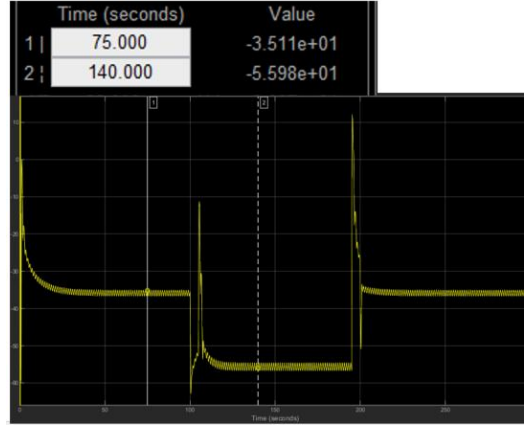

Fig. 9: Estimate of Thevenin impedance's angle.

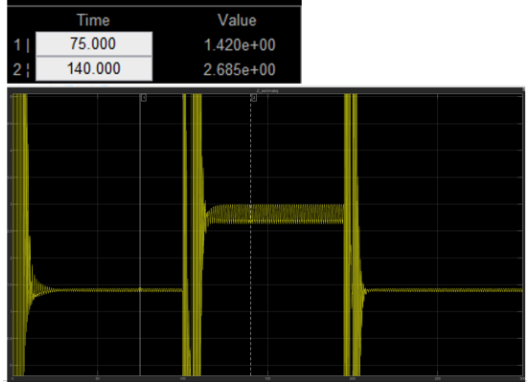

Fig. 10: Estimate of Thevenin impedance's magnitude.

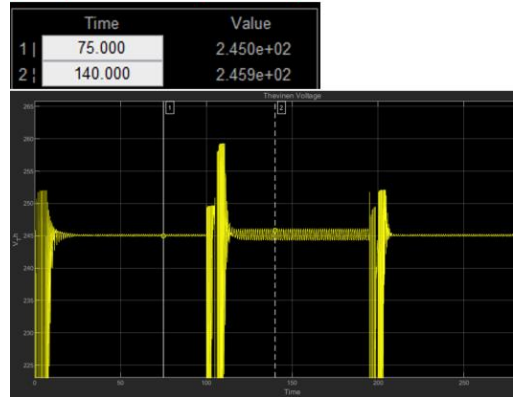

Fig. 11: Estimate of Thevenin voltage's magnitude.

# Appendix
## Bayesian Filtering
In a Bayesian approach, we are interested in developing the posterior probability of density function (pdf) with prior knowledge. Suppose we are interested in measuring $x_k$ at time k.
Let $x_k$, $k \in N$, be the state sequence:
$$x_k = f_k(x_{k-1}, u_{k-1}, v_{k-1}) \quad (1)$$
where $f_k$ is generally non-linear function of the previous state $x_{k-1} \in R^n$, $v_{k-1} \in N^n$ is state noise, $u_{k-1} \in R^n$ is known input.
Next, let $z_k \in R^n$ be the measurement function:
$$z_k = h_k(x_k, n_k) \quad (2)$$
Where $h_k$ is in generally non-linear measurement's function, $n_k \in N^n$. It is assumed that the initial PDF $p(x_0|z_0) \equiv p(x_0)$ is available. Using the system model at the k-time step, the prior PDF of the state can be determined,
$$p(x_k|z_{1:k-1}) = \int p(x_k|x_{k-1}) p(x_{k-1}|z_{1:k-1}) \, dx_{k-1} \quad (3)$$

The measurement $z_k$ is available at each time step k. so it can be used to update the prior. Using Bayes' rule, we obtain:
$$p(x_k|z_{1:k}) = \frac{p(z_k|x_k) p(x_k|z_{1:k-1})}{p(z_k|z_{1:k-1})} \quad (4)$$
where the normalizing constant is
$$p(z_k|z_{1:k-1}) = \int p(z_k|x_k) p(x_k|z_{1:k-1}) \, dx_k \quad (5)$$

## Kalman Filter
Kalman filter is a commonly used tool in statistical signal processing. In a state model assuming the posterior density in time k-1, $p(x_{k-1}|z_{k-1})$ and $p(x_k|z_k)$ are Gaussian. The random variables of $v_{k-1}$ and $n_k$ are independent and have probability distribution with covariances labeled ascombine pdf $Q_{k-1}$ and $R_k$. The state sequence $x_k$ and the measurement $z_k$ are linear functions. Hence the optimal Bayesian solution of equations (1) and (2) can be written as,
$$X_k = F_k x_{k-1} + B_k u_k + v_{k-1} \quad (6)$$
$$Z_k = H_k x_k + n_k \quad (7)$$
where $F_k$ and $H_k$ are matrices defining the linear function, these matrices, and the covariance matrices $Q_{k-1}$ and $R_k$ might change with each time step or measurement. Kalman filter is based on Bayesian filtering and can work in the following two phases.

### Predict stage:
The prediction stage can be described by the following two equations:
$$\hat{x}_{k|k-1} = F_k \hat{x}_{k-1|k-1} + B_{k-1} u_{k-1} \quad (6)$$
here $\hat{x}_{k|k-}$ is the estimate of the state at time k given observations up to time k and
$$P_{k|k-1} = F_k P_{k-1|k-1} + F_k^T + Q_{k-1} \quad (7)$$
where $P_{k|k-1}$ is the error covariance matrix.
### Update stage
The update stage can be described by the following equations:
$$\tilde{y} = Z_k + F_k \hat{x}_{k-1|k-1} \quad (8)$$
Where $\tilde{y}$ is innovation term,

$S_k = H_k P_{k|k-1} H_k^T + R_k$ (9)

Where $S_k$ is innovation covariance, and $R_k$ is the covariance of $n_k$,

$K_k = P_{k|k-1} H_k^T S_k^{-1}$ (10)

Where $K_k$ is Kalman gain,

$\hat{x}_{k|k} = \hat{x}_{k|k-1} + K_k \tilde{y}_k$ (11)

Is updated state estimate and

$P_{k|k} = (I - K_k H_k) P_{k|k-1}$ (12)

Is updated estimate covariance.

1. **Data Processing:**

Data loading, e.g., signals, and measurements.

2. **KF Initialization:**

Define measurement and noise covariances (Q and R)

Initialize estimate parameter, x_est = 0, and covariance, p = 0.

3. **KF Loop:**

Time step from 2 to N

**3.1 Prediction**

3.1.a Using the previous state, predict the next state (x_pred)

3.1.b Considering previous covariance and previous noise covariance, predict the covariance (P_pred)

**3.2 Update step**

3.2.a Considering measurement (z(k)) and its uncertainty, calculate Kalman gain (k)

3.2.b Based on previous information (Kalman gain, previous estimate, and measurement) update the estimate (x_est)

3.2.c Based on Kalman gain update previous covariance (P)

3.2.d Store current parameter estimate (x_est)

**4. Bayesian Updating:**

Computing the mean of all parameters calculates the final estimated parameter.

**5. Displaying the result and Plotting:**

### Comparison of RWLS and Kalman filtering in estimating Thevinen Impedance Values

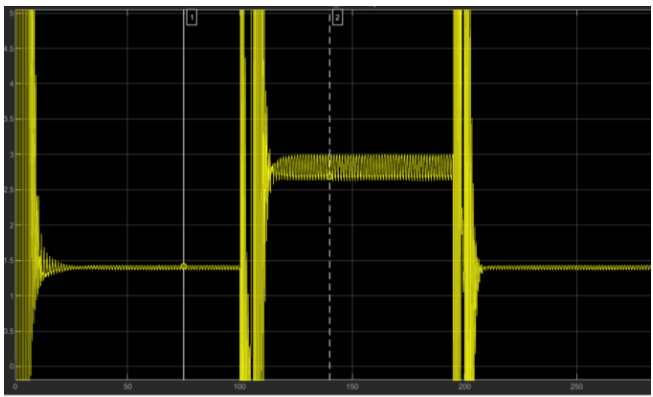
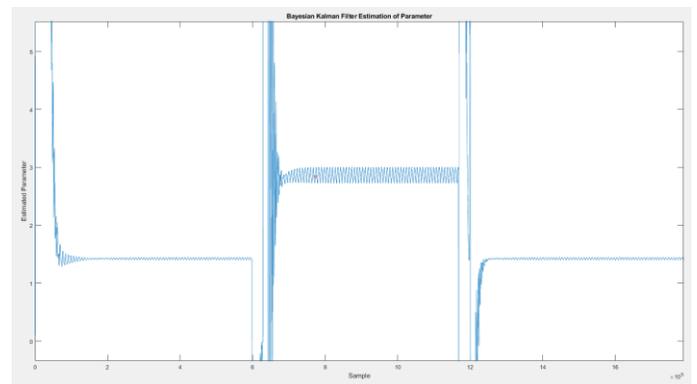

$Z^{th}$ using RWLS                     $Z^{th}$ using Kalman Filter